# Role of Dirac cones in magnetotransport properties of REFeAsO (RE=rare earth) oxypnictides

I.Pallecchi [1], F.Bernardini [2], F.Caglieris [1], A.Palenzona [1,3], S.Massidda [2], M.Putti [1,4]

[1] CNR-SPIN-Genova corso Perrone 24, 16152, Genova, Italy
[2] CNR-IOM and Physics Department, University of Cagliari, 09042 Monserrato (CA), Italy
[3] Chemistry and Industrial Chemistry Department, University of Genova, via Dodecaneso 31, 16146 Genova, Italy
[4] Physics Department, University of Genova, via Dodecaneso 33, 16146 Genova, Italy

**Abstract**
In this work we study the effect of the rare earth element in iron oxypnictides of composition REFeAsO (RE=rare earth). On one hand we carry out Density Functional Theory calculations of the band structure, which evidence the multiband character of these compounds and the presence of Dirac cones along the Y-Γ and Z-R directions of the reciprocal space. On the other hand, we explore transport behavior by means of resistivity, Hall resistance and magnetoresistance measurements, which confirm the dominant role of Dirac cones. By combining our theoretical and experimental approaches, we extract information on effective masses, scattering rates and Fermi velocities for different rare earth elements.

**Introduction**
Since the discovery of superconductivity in doped LaFeAsO [1], it has been clear that the understanding of unconventional superconducting mechanisms in iron oxypnictides must necessarily include the investigation of the parent compounds, which all exhibit a spin density wave (SDW) ground state below a certain temperature $T_{SDW}$. It has been found that superconductivity appears and SDW disappears upon doping in several other members of the REFeAsO family (RE=rare earth), among which doped LaFeAsO [1], CeFeAsO [2], PrFeAsO [3], SmFeAsO [4], with different critical temperatures ranging from 26K to 56K. In this context, the role of the rare earth element in the parent compound could yield interesting information. From the experimental standpoint, the influence of the rare earth on the structural [5,6], magnetic [7,8,9] and electronic properties [5] has been explored. It has been found that the SDW transition temperatures vary little with rare earth, while the Fe magnetic moment $\mu_{Fe}$ varies considerably, even if the experimental results on Fe moment are somewhat scattered and contradictory [9] (recent values $\mu_{Fe}$= 0.36$\mu_B$, 0.8 $\mu_B$, 0.35 $\mu_B$, 0.34 $\mu_B$ have been measured in LaFeAsO [10], CeFeAsO [11], PrFeAsO [12] and SmFeAsO [13], respectively, above the ordering temperature of Ce and Pr magnetic moments). The transport properties are qualitatively similar among the compounds, however the relative contributions of multiple carrier bands to transport properties undergo systematic changes with varying rare earth [5]. Also the sensitivity of $T_{SDW}$ and Fe magnetic moment to external pressure has been found to be rare earth dependent, being stronger for RE=La and weaker for RE=Sm [7]. Early first principles investigation on the effect of the rare earth on the band structure [14] have pointed out the absence of significant changes in the band dispersion that could justify the large differences in the $T_c$ values observed in doped oxypnictides. Accurate calculations [15] including correlation effects have shown that dispersion at Fermi level is not affected by the RE while the RE dependent degree of hybridization of *4f* electrons with *3d* Fe is likely the origin of the different low temperature magnetic behavior of REFaAsO.

In this work, we undertake a theoretical as well as experimental investigation of four REFeAsO compounds (RE=Ce, La, Pr or Sm), in order to identify common features and differences related to the effect of the chemical composition on the band structure and magnetotransport properties. At variance, with the investigation of Ref. 14 we explicitly include semi-core states as valence ones and consider correlation effects beyond the local density approximation (LDA) by the Hubbard U term. Our outcomes evidence the similarity of the band dispersion within the REFeAsO family and the formation of very similar Dirac cone structures close to the Fermi level. Consistently, our experimental magnetotransport data are qualitatively similar for the different compounds and confirm the presence of Dirac cones.

**Band structure calculation**
We compute the REFeAsO (RE=Ce, La, Pr or Sm) electronic band structure in the framework of Density Functional Theory and Generalized Gradient approximation of Perdew, Burke and Ernzerhof [16] as implemented in the Wien2K full-potential APW+lo package [17]. Since Ce, Pr and Sm 4$f$ shells are not empty, we add an on-site Coulomb correlation contribution in the spirit of DFT+U [18]. The U parameter is set to 9.7 eV for Ce and Sm, to 7.6 eV for Pr [19,20]. The muffin-tin radii for Fe, As and O are chosen equal to 2.3, 2.1, 1.9 Bohr, respectively. Rare earth radii are 2.3 Bohr for La, Ce and Pr, 2.4 Bohr for Sm. The plane-wave cutoff is chosen to be $R_{mt} \times K_{max}=9$. Semicore Fe 3$s$ and 3$p$, rare-earth 4$d$ states are explicitly taken into account as valence states. Magnetism is treated within the collinear formalism. To represent the correct stripe phase we use a cell with four formula units (Pccm, space group No. 49). We use experimental values for the structural parameters that are listed in Table I and III. The comparison of the band structure computed using the experimental positions for LaFeAsO with the band structure obtained using the optimized structure shows that for the magnetic phase the optimization of the internal structural parameters does not change the band structure across $E_F$. Therefore, in this work we use the experimental structural parameters available in literature for all the calculations. We point out that for the Ce, Pr and Sm compounds, our calculations assumes the U parameter, which makes our calculations no more strictly *ab-initio*. Therefore, a structural optimization in this framework would be less reliable and further justifies our choice of experimental structural parameters. Spin orbit corrections to the band structure are small and we neglect them in the following. As usual for DFT, we find magnetic moments on Fe to be overestimated with respect to experimental values, being 2.13, 2.08, 2.09, 2.02 Bohr magnetons for La, Ce, Pr and Sm compounds respectively [21].

In agreement with our recent investigation on LaFeAsO [22], we find that Dirac cones form near the Fermi level. Cones are absent in the non-magnetic phase, but they appear in the magnetic stripe configuration. Indeed, a gap opens along the direction of antiferromagnetic ordering for the stripe phase (X-Γ) leaving a contact point (the Dirac cone) along the orthogonal Y-Γ direction. As in our previous work [22], we compute the band structure, assuming the magnetic moments $\mu_{Fe}$ obtained by the DFT calculation, larger than most of the literature experimental values, which in turns are widely scattered and dependent on the experimental method [9,13]. In Table II we summarize some of the outcomes of our calculations, relevant to the experimental data analysis presented hereafter. In Fig. 1(a) we show the band structure dispersions of the REFeAsO compounds along the Y-Γ and Z-R directions as obtained by our calculations, with the Fermi levels positioned so as to yield exact electron-hole compensation, as expected for stoichiometric compounds. In Fig. 1(b) we show the CeFeAsO compound Fermi surface (FS). Keeping in mind that Y-Γ is the direction of ferromagnetic ordering for the stripe phase, in Fig 1(a) we see that the bands cross along the Y-Γ and Z-R directions, forming cones midway. The choice of assuming the $\mu_{Fe}$ predicted by DFT yields quite deep cone vertices. These cones are shown as two blue cylindrical FS sheets in Fig 1(b). A hole pocket is found along the Γ-Z direction in Fig 1(a) which corresponds to the red cylindrical FS at the Brillouin zone centre in Fig 1(b). Holes show a small dispersion along Γ-Z in the La and Sm compounds only. As for the cones, the dispersion along $k_z$ at $E_F$ is virtually negligible for all the compounds. The shape of the FSs in Fig. 1(b) confirms this finding. A small bending of the cone FSs is the only dispersion effect along the out-of-plane direction. This indicates that the cone vertices are closer to Γ at $k_z=0$ than they are to Z at $k_z=\pm\pi/c$. The sizes and shapes of the FS cylinder sections remain almost unchanged along $k_z$. We find instead a different behavior for the in-plane dispersion as evidenced along the Y-Γ and Z-R directions in Fig. 1(a). The lattice compression along the basal plane increases the bandwidth. This is confirmed by the trend in Table II toward higher Fermi velocities in the compounds with La, Ce, Pr, Sm, respectively.

|  | $z_{RE}$ | $z_{As}$ | Reference |
|---|---|---|---|
| **LaFeAsO** | 0.1417 | 0.6507 | 23 |
| **CeFeAsO** | 0.1413 | 0.6546 | 24 |
| **PrFeAsO** | 0.1399 | 0.6565 | 25 |
| **SmFeAsO** | 0.1368 | 0.6609 | 26 |

**Table I:** Experimental values of the REFeAsO internal structure parameters used in the calculations.

In a stoichiometric compound holes and electrons should be compensated ($n_e=n_h$). This is contrary to our experimental evidence based on the interpretation of Hall effect and magnetoresistance data as discussed in the next section. Therefore here we do not use the Fermi energy obtained from the DFT calculation to estimate the carriers densities. Instead we estimate the Fermi energy so as to match the experimental carriers densities for electrons and holes and then use band structure parameters averaged over the Fermi surface, such as Fermi velocities and effective masses (reported in Table II), to extract further information from the experimental data, as discussed in the next section.

We further comment on the discrepancy between calculated and experimental carrier densities by referring to a recent work by Fanfarillo et al.[27], where it is pointed out that in iron pnictides correlation can be large enough to produce deviations from standard Boltzmann theory. This may lead to large values for the Hall coefficient even in compensated materials, that cannot be simply recast in a renormalization of the scattering times, as it induces a mixing of electron and hole currents and yields the experimentally observed enhancement of $|R_H|$. This could argument could explain the discrepancy between calculated and measured carrier densities. Another possible reason is the uncertainty on the Hubbard term U as an input parameter of the calculation.

We also investigate the effect of $\mu_{Fe}$ on the most relevant band structure parameters (effetive masses, Fermi velocity) performing band structure calculations at different $\mu_{Fe}$ values. In these calculations, we use a potential obtained as the interpolation of two potentials, namely the self-consistent potential coming from a non magnetic (spinless) calculation and the self-consistent potential from the magnetic system in the stripe phase, in the spirit of the virtual crystal approximation:

$$V(\mu'_{Fe}) = (\mu'_{Fe}/\mu_{Fe})V(\mu_{Fe}) + (1-\mu'_{Fe}/\mu_{Fe})V(0) \qquad (1)$$

where $V(\mu'_{Fe})$ is the interpolated potential, $V(\mu_{Fe})$ the potential in the stripe phase and $V(0)$ the potential obtained in a non magnetic DFT calculation.

Lowering $\mu_{Fe}$ down to 0.8 $\mu_B$ yields a 30% reduction of the Fermi velocity and increases the electron band dispersion along the $k_z$. By further lowering $\mu_{Fe}$, we find that an electron pocket with parabolic dispersion appears along the X-Γ direction in all the four REFeAsO compounds and the cone vertices shift closer to the Fermi level. Eventually, when we assume $\mu_{Fe}$ values similar to the experimental ones, the band gap closes leading to the disappearance of cones. From the comparison with experimental data, presented in the following section, we gather that the best band structure description of these compounds is obtained with $\mu_{Fe}$ values above 0.8 $\mu_B$, yielding presence of SDW gap and Dirac cones and absence of any parabolic electron-type bands crossing the Fermi level. On the whole, the band structure evolution with changes of $\mu_B$ does not change dramatically the band structure parameters that we use for the interpretation of experimental data. Therefore, in the following analysis and discussion of experimental data, we use the results reported in Table II, obtained by the DFT calculation with theoretical values of $\mu_{Fe}$ and with the Fermi levels placed in such a way to match the carrier densities extracted from our magnetotransport data, as described in the following section.

| | $v_F$ (Km/s) | $m_{DC}/m_0$ | $m_h/m_0$ |
|---|---|---|---|
| **LaFeAsO** | 157 | 0.017 | 0.24 |
| **CeFeAsO** | 136 | 0.046 | 0.29 |
| **PrFeAsO** | 187 | 0.013 | 0.27 |
| **SmFeAsO** | 205 | 0.026 | 0.49 |

**Table II:** Band structure parameters averaged over the whole Fermi surface, obtained from DFT calculations: Dirac cone Fermi velocities, Dirac cone effective masses and in-plane component of hole effective masses in units of free electron mass $m_0$. These parameters are obtained with the Fermi levels placed in such a way to match the carrier densities extracted from our magnetotransport data, as described in the following section.

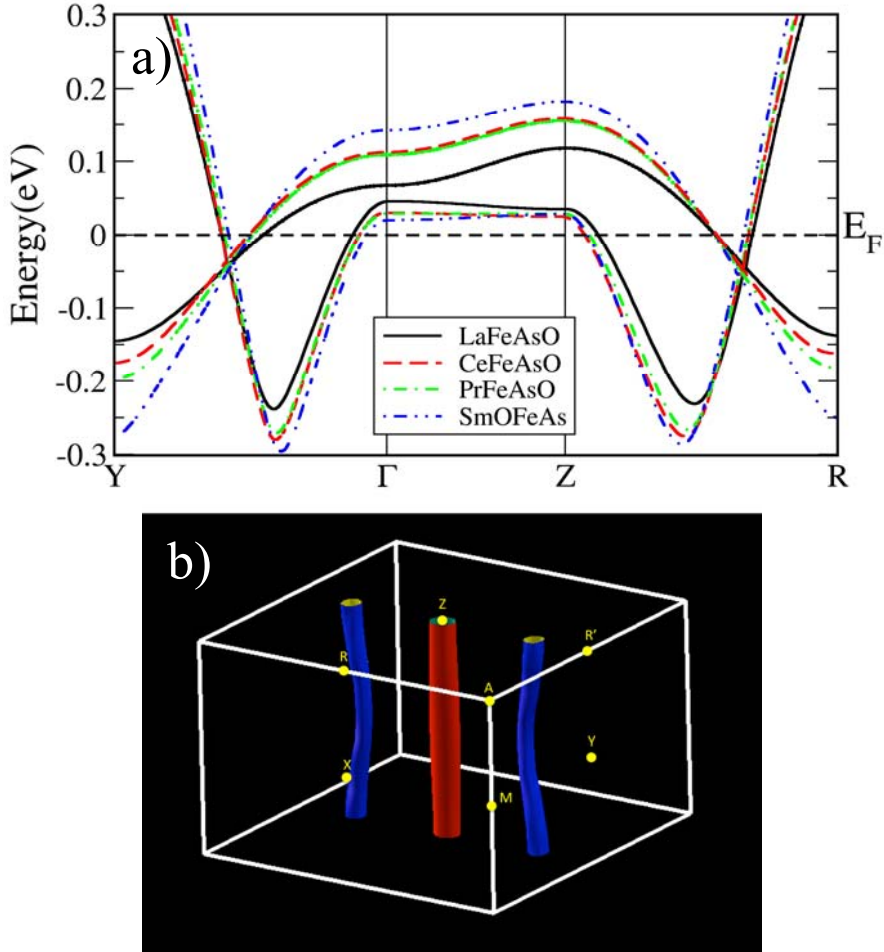

**Figure 1:** (color online) Upper panel: calculated band structure dispersion of REFeAsO. Lower panel: Fermi surface of CeFeAsO. Red (blue) cylinders are holes (electrons). For the other compounds the shape of the Fermi surface is similar. In these figures, the Fermi level is positioned as expected for stoichiometric compounds, yielding exact electron-hole compensation.

**Experimental results and discussion**
REFeAsO (RE=Ce, La, Pr or Sm) samples are prepared by solid state reaction at ambient pressure from RE, As, Fe and $Fe_2O_3$. REAs are first synthesized from pure elements in an evacuated and sealed glass tube at a maximum temperature of 550°C. Successively, the samples are obtained by mixing the REAs, Fe and $Fe_2O_3$ powders in stoichiometric proportions, uniaxial pressing to turn the powders into pellets, heat treating the pellets in an evacuated and sealed glass tube at 1000-1100 °C for 24 hours and finally furnace cooling. The structures of the samples are analyzed by X-ray diffraction in a Guinier camera, with Si as internal standard. The X-ray patterns evidence the presence of a single phase with only weak extra peaks of secondary phases at low angle. The lattice parameters of the tetragonal cell containing two Fe ions are summarized in Table III.

|  | a (Å) | c (Å) |
|---|---|---|
| LaFeAsO | 4.039(3) | 8.751(2) |
| CeFeAsO | 4.001(2) | 8.655(2) |
| PrFeAsO | 3.985(1) | 8.595(2) |
| SmFeAsO | 3.940(1) | 8.502(2) |

**Table III:** Structural parameters of the tetragonal cell containing two Fe ions measured by X-ray diffraction on the four REFeAsO samples. The figures in parentheses indicate the uncertainty on the last digit. Note that in the *ab initio* calculations and in the fitting of experimental data we consider the unit cell containing four Fe ions, with in-plane lattice parameter a' rotated by 45°( a'=a·√2).

Magnetotransport behavior is measured in a Physical Properties Measurement System (PPMS) by Quantum Design at temperature from room temperatures down to 2K and in magnetic fields up to 9T.

In Fig. 2 we present the resistivity curves of the series of REFeAsO samples with different rare earths La, Pr, Ce and Sm. At high temperature the curves are weakly temperature dependent and in all cases the resistivity values are in the range of few mΩ·cm. SmFeAsO has the lowest resistivity, while LaFeAsO has the largest one. An abrupt change of regime is observed around the structural transition. Indeed, about 10 K below the structural transition, a curvature change marks the magnetic transition temperature $T_{SDW}$, below which the spin density wave (SDW) ground state is established. The $T_{SDW}$ values, determined from the maximum of the first derivative, are all in the temperature window between 130K and 145K (see Table IV). At lower temperatures the curves exhibit metallic behavior. The LaFeAsO sample eventually shows a resistivity upturn below 30K, likely caused by weak localization (or antilocalization).

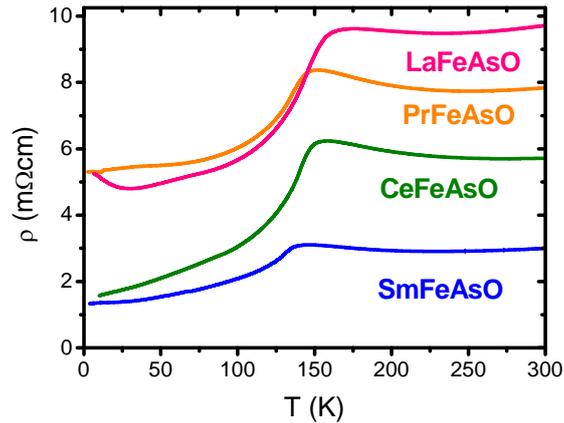

**Figure 2:** (color online) resistivity curves of the REFeAsO samples.

|  | $T_{SDW}$ (K) | ρ at 5 K (mΩ cm) | ρ at 20 K (mΩ cm) | ρ at 300 K (mΩ cm) | $R_H$ at 5K ($m^3$/C) | $R_H$ at 20K ($m^3$/C) |
|---|---|---|---|---|---|---|
| LaFeAsO | 145 | 5.35 | 4.88 | 9.76 | -3.98·$10^7$ | -3.20·$10^7$ |
| CeFeAsO | 140 | 1.49 | 1.73 | 5.71 | -1.29·$10^7$ | -1.26·$10^7$ |
| PrFeAsO | 139 | 5.31 | 5.39 | 7.84 | -2.15·$10^7$ | -1.89·$10^7$ |
| **SmFeAsO** | 130 | 1.34 | 1.38 | 2.98 | -1.86·10⁷ | -1.74·10⁷ |

**Table IV:** List of parameters of the series of REFeAsO samples: spin density wave transition temperatures, resistivity and Hall resistance values at selected temperatures. The values at 20K are reported, because the set of fitting parameters at this temperature is used as a reference for Fermi level positioning in *ab initio* band calculations.

In Fig. 3 we present Hall resistance $R_H$ curves as a function of temperature. Above $T_{SDW}$ all the $R_H$ values are very small (~$10^{-8}$ m³/C or smaller) and weakly temperature dependent, suggesting that in this regime electron-type and hole-type carriers are virtually compensated and give rise to a vanishing small $R_H$. Below $T_{SDW}$, similarly for all the samples, $R_H$ curves increase in magnitude and are negative in sign. This suggests (i) that a carrier condensation occurs in correspondence of the opening of the SDW gap and of the Fermi surface reconstruction and (ii) that at low temperature transport is dominated by electron type carriers. Additionally, we can infer that below $T_{SDW}$ the simultaneous increase in |$R_H$| and decrease in ρ is accounted for in terms of significant increase in electron mobility in the SDW regime.

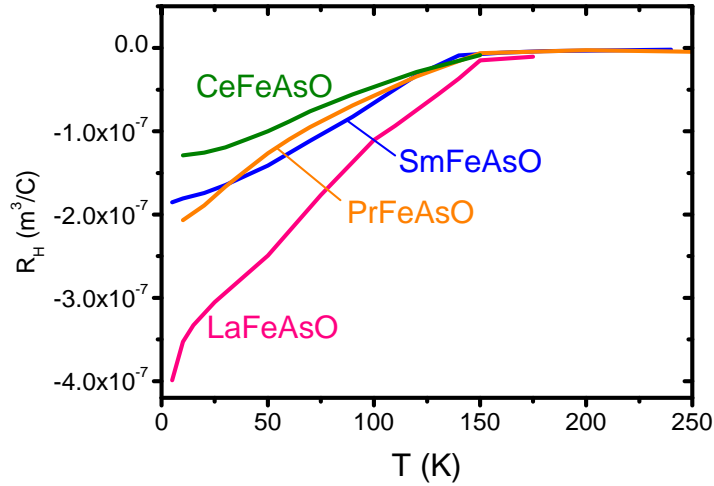

**Figure 3:** (color online) Hall resistance curves of the REFeAsO samples

In Fig. 4 we finally present the magnetoresistance $\Delta\rho/\rho(0)=(\rho(H)-\rho(H=0))/\rho(H=0)$ curves of the series of samples as a function of applied perpendicular field at different temperatures. In all cases, the magnetoresistance at high fields ($\mu_0 H$ larger than few Tesla) exhibits a linear behavior. This has been considered a hallmark of the presence of Dirac cones in the band structure at the Fermi level in iron pnictides [28,29,30,31,32,33,22], well explained by the model of quantum magnetoresistance developed by Abrikosov [34,35]. This linear magnetoresistance contribution is superimposed to a semiclassical cyclotron contribution, which obeys the usual $H^2$ dependence at low fields. Moreover, at low temperatures and low fields further additional magnetoresistance contributions, different for each sample, are present. Namely, below 5K the CeFeAsO sample exhibits humps related to the ordering of the Ce magnetic moments [36], while the LaFeAsO sample shows well visible dips at $\mu_0 H<1T$. The latter feature, also weakly seen in the PrFeAsO sample, as emphasized in the inset of Fig. 4, could be accounted for in terms of rapidly saturating cyclotron magnetoresistance, in presence of huge mobility values [22], however it seems to be more realistically due to the weak antilocalization mechanism [37,38,39,40,41] assuming a phase coherence length of about 400 nm. The latter explanation is consistent with the two-dimensional character of these iron pnictides, akin to topological insulators [38,39,40,41], as well as with the observation of the resistivity upturn at low temperature (compare Fig. 2).

In order to extract information on band structure and transport parameters of our samples, we try to fit magnetotransport data, assuming a two band model with hole and electron contributions. We focus on data above 20K for all the samples, so that we can consider only the linear magnetoresistance and the cyclotron terms. For the magnetoresistance, Hall resistance and resistivity we use the following formulas, as in ref. [22,42]:

$$\frac{\rho(H)-\rho(0)}{\rho(0)} = \frac{\left(\sum_i^N \frac{1}{\left(\gamma_i \mu_0 |H| + \frac{1}{\sigma_i}\right)\left(1+(\mu_i \mu_0 H)^2\right)}\right)}{\left[\left(\sum_i^N \frac{1}{\left(\gamma_i \mu_0 |H| + \frac{1}{\sigma_i}\right)\left(1+(\mu_i \mu_0 H)^2\right)}\right)^2 + \left(\sum_i^N \frac{\mu_i \mu_0 H}{\left(\gamma_i \mu_0 |H| + \frac{1}{\sigma_i}\right)\left(1+(\mu_i \mu_0 H)^2\right)}\right)^2\right] \frac{1}{\sum_i^N \sigma_i}} - 1 \quad (2.a)$$

$$R_H = \frac{(\mu_1 \sigma_1 + \mu_2 \sigma_2) + \mu_2 \mu_1 (\mu_2 \sigma_1 + \mu_1 \sigma_2)(\mu_0 H)^2}{(\sigma_1 + \sigma_2)^2 + (\mu_2 \sigma_1 + \mu_1 \sigma_2)^2 (\mu_0 H)^2} \quad (2.b)$$

$$\rho = \left(\frac{1}{\sigma_1 + \sigma_2}\right) \quad (3.b)$$

where $\mu_0$ is the vacuum magnetic permeability, N=2 is the number of bands, $\mu_i$ and $\sigma_i$ are the mobility and conductivity of the *i-th* band and $\gamma_i$ is the coefficient of the linear magnetoresistance, so that $\gamma_i$=0 if the *i-th* band is parabolic and $\gamma_i$≠0 if the *i-th* band has linear dispersion relation. We leave as free parameters the carrier densities and mobilities of each band, as well as the linear magnetoresistance coefficients. We fit simultaneously the $R_H$ and resistivity values, as well as the magnetoresistance curves. We note that in order to reproduce the shape of the magnetoresistance curves and in particular the linear trend at high fields, the cyclotron term must deviate from the $H^2$ behavior and change curvature at fields of few Tesla, which is equivalent to fixing two constraints. The fitting magnetoresistance curves and the separate cyclotron and linear contributions are displayed in Fig. 5. In the end, even though the solution is not univocal from a mathematical point of view, we identify a unique solution for each sample which is physically plausible. This corresponds to the presence of an electron type band having higher mobility and lower carrier concentration plus a hole type band having lower mobility and higher carrier concentration. The fitting parameters at 20K are reported in Table V and their behavior as a function of temperature is displayed in Fig. 6. In the upper panel, it can be seen that for all the compounds the electron carrier density $n_e$ is two orders of magnitude smaller than the hole carrier density $n_h$, with $n_e$ in the range $5 \cdot 10^{-5}$-$3 \cdot 10^{-4}$ per unit cell and $n_h$ in the range $2 \cdot 10^{-3}$-$3 \cdot 10^{-3}$ per unit cell. The SmFeAsO sample has the largest carrier density, consistent with its lower resistivity. Both the carrier densities vary very weakly with temperature at low temperature, but as $T_{SDW}$ is approached the hole density increases, while the electron density decreases. Also the carrier mobilities in the middle panel of Fig. 6 show a very similar behavior for all the compounds. The electron mobilities $\mu_e$ are one order of magnitude larger than the hole mobilities $\mu_h$, with $\mu_e$ around 0.1 m$^2$V$^{-1}$s$^{-1}$ and $\mu_h$ in the range $4 \cdot 10^{-3}$-$2 \cdot 10^{-2}$ m$^2$V$^{-1}$s$^{-1}$ at low temperature. Both hole and electron mobilities decrease with increasing temperature. We identify the high mobility electron band of each compound with the Dirac cone band and the low mobility hole band with the parabolic dispersion hole band predicted by *ab initio* calculations. The mobility decrease with increasing temperature is reasonably understood in terms of increased scattering with phonons and spin fluctuations. Also the temperature dependence of the carrier densities is consistent with expectations. Indeed, at $T_{SDW}$ the spin density wave gap opens up and the holes start to localize into the antiferromagnetic ordered state, so that $n_h$ decreases with decreasing temperature below $T_{SDW}$ and saturates at a constant value at low temperature. On the other hand, at $T_{SDW}$ the Dirac cones are formed and electrons start to populate the cone states, so that $n_e$ increases with decreasing temperature. Hence, below $T_{SDW}$ transport is increasingly dominated by the Dirac electron band with decreasing temperature, as demonstrated by the experimental evidences of negative $R_H$ and decrease of resistivity despite the simultaneous condensation of carrier density. In particular, this decrease of resistivity despite the simultaneous condensation of carrier density below $T_{SDW}$ is explained within a two-band picture in terms of the corresponding significant increase of average mobility, increasingly dominated by $\mu_e$ much larger than $\mu_h$.

Regarding the cone population, the CeFeAsO and SmFeAsO samples have larger $n_e$, while LaFeAsO and PrFeAsO samples have smaller $n_e$, indicating that in the latter samples the Fermi level is closer to the Dirac cone vertex. For all the compounds, the carrier density values in the range $5 \cdot 10^{-5}$-$3 \cdot 10^{-4}$ per unit cell should fill only the lowest Landau level (LL) in the Dirac cones at a field of 9T (the number $N_{LL}$ of filled LL can be estimated as $N_{LL} = n_{2D} 2\pi\hbar/q\mu_0 H$, where q is the electron charge and $n_{2D}$ is the carrier density per unit area in the cone states). However, in the case of CeFeAsO and SmFeAsO, more than one LL is populated at fields lower than 3-4T. This is not in contradiction with the linear magnetoresistance contribution, visible also at low fields. Indeed, although Abrikosov model for linear quantum magnetoresistance [34,35] has been developed in the limit of a single occupied LL, linear magnetoresistance is also predicted and observed in the situation of several occupied LLs [43,44].

As mentioned above, we can combine information from band structure calculations and from experimental data, by positioning the Fermi level in such a way to match the values of the carriers densities obtained by the data fit. Once fixed these energy values, we estimate Fermi velocities and effective masses averaged over the Fermi surface and report the corresponding values in Table II. Using these values plus the carrier mobilities obtained from data fittings, the scattering times $\tau_e$ and $\tau_h$ for electrons and holes and are easily obtained. For example, LaFeAsO and PrFeAsO have the lowest scattering times among these samples for Dirac electrons, namely $\tau^{(e)}=9.4 \cdot 10^{-15}$s and $\tau^{(e)}=6.5 \cdot 10^{-15}$s, respectively, while SmFeAsO and CeFeAsO have slightly larger values $\tau^{(e)}=1.4 \cdot 10^{-14}$s and $\tau^{(e)}=2.6 \cdot 10^{-14}$s. As for the parabolic hole scattering times we find the smallest $\tau^{(h)}=9.7 \cdot 10^{-15}$s in SmFeAsO, the largest $\tau^{(h)}=3.8 \cdot 10^{-14}$s in CeFeAsO and similar intermediate values $\tau^{(h)}=1.3 \cdot 10^{-14}$s and $\tau^{(h)}=1.4 \cdot 10^{-14}$s in LaFeAsO and PrFeAsO, respectively. On the whole, hole and electron scattering rates are comparable in all the compounds and the large difference in electron and hole mobilities, about one order of magnitude, is explained by the difference in their effective masses. Using the calculated Fermi velocities in Dirac cones reported in Table II, we estimate a Dirac electron mean free paths of 1.5nm, 3.6nm, 1.2nm and 2.9nm for LaFeAsO, CeFeAsO, PrFeAsO and SmFeAsO, respectively. Clearly, an effective mass renormalization due to correlation effects would enhance these scattering rate values.

In the lower panel of Fig. 6 the fitting parameters $\gamma_e$ of the linear magnetoresistance contributions are plotted and their values at 20K are reposted in Table V. It can be seen that $\gamma_e$ low temperature values are in the range $2 \cdot 10^{-6}$-$9 \cdot 10^{-6}$ $T^{-1}$. The largest $\gamma_e$ is observed in SmFeAsO, while the smallest one in CeFeAsO. We point out that these values in the range $10^{-6}$ $T^{-1}$ come out from experimental slopes in the range 0.0006-0.007 $T^{-1}$ (indicated in round parentheses in Table V) when band conductivities are summed up as explained in [42], where eq. (2.a) is derived. The $\gamma_e$ values can be directly compared with the predictions of Abrikosov model [34]. Indeed, from a microscopic point of view, we can relate these values with the expression:

$$\rho = \frac{1}{2\pi} \left( \frac{e^2}{\varepsilon_\infty \hbar v_F} \right)^2 \frac{N_i}{en^2} \mu_0 H \ln(\varepsilon_\infty)$$

(3)

where $v_F$ is the Fermi velocity at the Dirac cones, $\varepsilon_\infty$ is the high frequency dielectric constant and $N_i$ is the impurity concentration. From our DFT calculation we extract the Fermi velocity values in the Dirac cones, averaged over the whole cone related Fermi surface, for the four compounds. The values range between $v_F$=136 Km/s for CeFeAsO and $v_F$=205 Km/s for SmFeAsO. It comes out that these values are not inversely correlated to the linear magnetoresistance contributions reported in Table V for the four samples, even if the different carrier densities are kept into account. Clearly, in eq. (3) the unknown impurity concentration $N_i$ cannot be assumed to be the same for all the samples and this does not allow a direct comparison between calculated $v_F$ and experimental linear magnetoresistance contributions.

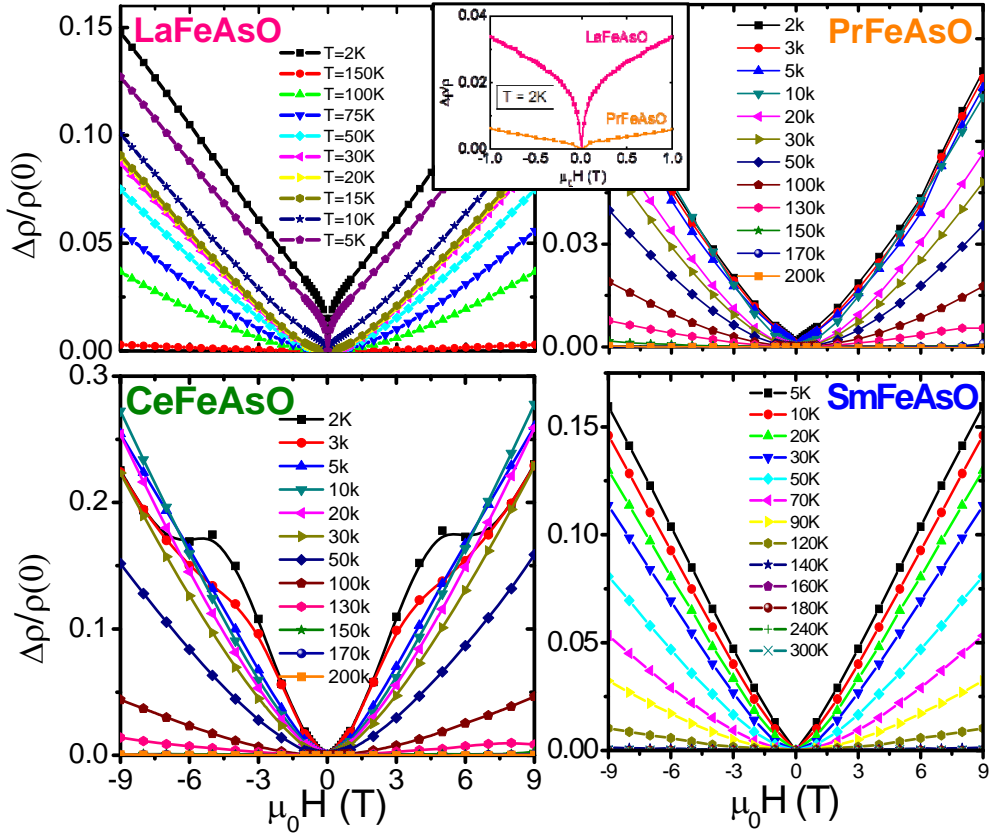

**Figure 4:** (color online) Magnetoresistance curves of the REFeAsO samples at different temperatures. In the inset, the magnetoresistance dip observed at low temperature and low field in LaFeAsO and PrFeAsO samples is zoomed.

|  | $n_e$ (carriers per unit cell) | $n_h$ (carriers per unit cell) | $\mu_e$ (m$^2$V$^{-1}$s$^{-1}$) | $\mu_h$ (m$^2$V$^{-1}$s$^{-1}$) | $\mu_e/\mu_h$ | $n_e/n_h$ | $\gamma_e$ (T$^{-1}$) |
|---|---|---|---|---|---|---|---|
| LaFeAsO | 5.44·10$^{-5}$ | 3.34·10$^{-3}$ | 0.0974 | 0.00915 | 10.6 | 0.0163 | 6·10$^{-6}$ (0.0018) |
| CeFeAsO | 3.00·10$^{-4}$ | 3.19·10$^{-3}$ | 0.101 | 0.0229 | 4.4 | 0.0940 | 2·10$^{-6}$ (0.0069) |
| PrFeAsO | 4.68·10$^{-5}$ | 3.04·10$^{-3}$ | 0.0872 | 0.00926 | 9.4 | 0.0154 | 6·10$^{-6}$ (0.0006) |
| SmFeAsO | 2.13·10$^{-4}$ | 3.06·10$^{-2}$ | 0.0966 | 0.00349 | 27.7 | 0.00696 | 4·10$^{-6}$ (0.007) |

**Table V:** List of fitting parameters for REFeAsO for experimental values of magnetoresistance, resistivity and $R_H$ at 20K, namely carrier densities (per unit cell, with unit cell containing four Fe ions) and mobilities of the two bands, ratios of the two band mobilities and carrier densities, linear magnetoresistance coefficients to be compared with eq. (3) (in round parentheses are the experimental slopes, as explained in the text).

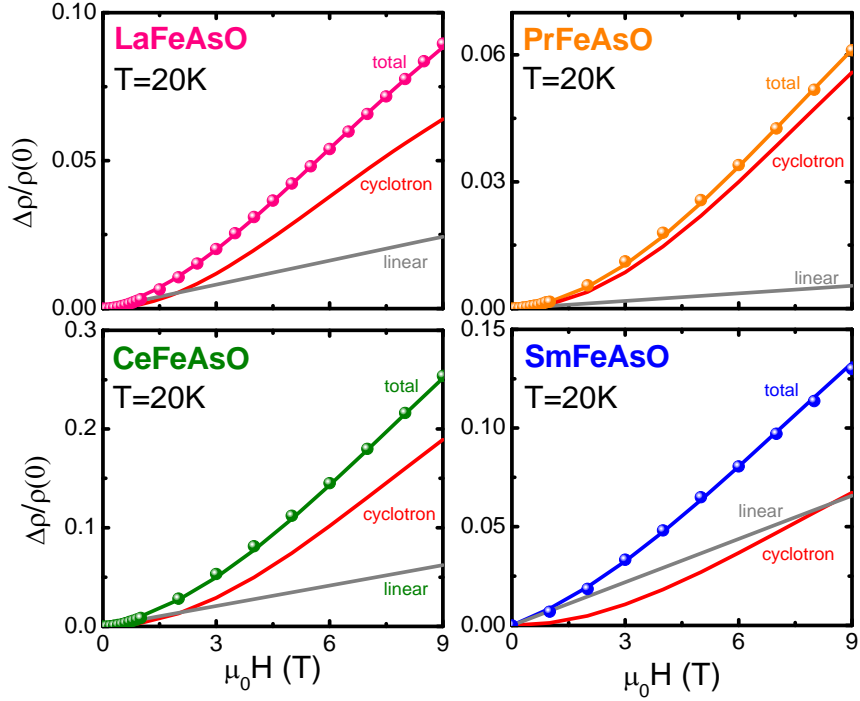

**Figure 5:** (color online) Experimental magnetoresistance curves of the REFeAsO samples at 20K (symbols). Also shown are the fitting curves of the total magnetoresistance and its separate cyclotron and linear contributions (continuous lines).

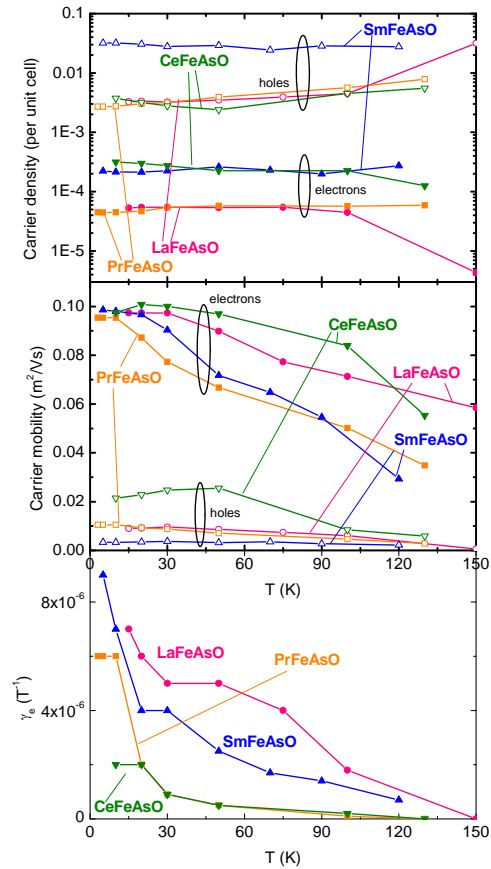

**Figure 6:** (color online) Temperature dependence of the fitting parameters for REFeAsO, namely carrier densities (upper panel, in logarithmic scale), mobilities (middle panel) and linear magnetoresistance coefficients (lower panel).

**Conclusions**

We investigate the role of the rare earth element in the band structure and magnetotransport properties of REFeAsO parent compounds. Resistivity curves show that the SDW transition temperatures vary at most by 10% among the compounds. Also Hall resistance curves have a similar temperature behavior, which suggests that a sharp decrease of carrier density occurs in correspondence of the opening of the SDW gap and of the Fermi surface reconstruction. Transport properties have a strongly multiband character, however at low temperature, transport appears to be dominated by electron type carriers. Thanks to the comparison with DFT results, we identify these electron bands with Dirac cones close to the Fermi level. Indeed, DFT calculations evidence the presence of Dirac cones along the Y-Γ and Z-R directions of the Brillouin zone for all the compounds. Linear magnetoresistance curves provide further evidence of the presence of Dirac cones. Our joined theoretical and experimental results do not converge in identifying consistently and unambiguously a clear trend of electronic parameters, such as effective masses and carrier densities, as a function of the rare earth. This could be due to (i) the strong influence of the Fe magnetic moment on the band structure which likely varies among the different compounds [9] but is not easily determined; (ii) correlation effects [27], (iii) possible extrinsic effects, such as slight accidental off-stoichiometry, impurities or microstructure, in our samples, which affect the measured transport properties, masking the weaker effect of the different rare earths; (iv) the fact that the rare earth type plays indeed a minor role in determining the band structure and the transport properties. Despite this open issue, a multiband fitting of magnetotransport data, joined with input parameters taken from DFT calculations, allows us to extract carrier densities and mobilities in the Dirac electron-type and parabolic hole-type bands. The low temperature carrier densities in the Dirac cones $n_e$ are in the range $5 \cdot 10^{-5}$-$3 \cdot 10^{-4}$ per unit cell and in parabolic bands $n_h$ in the range $2 \cdot 10^{-3}$-$3 \cdot 10^{-3}$ per unit cell, with a ratio $n_e/n_h$ around $10^{-2}$. The low temperature carrier mobilities in the Dirac cones $\mu_e$ are around $0.1$ $m^2V^{-1}s^{-1}$ and in parabolic bands $\mu_h$ in the range $4 \cdot 10^{-3}$-$2 \cdot 10^{-2}$ $m^2V^{-1}s^{-1}$, with a ratio $\mu_e/\mu_h$ around 10. Using DFT calculated effective masses, we estimate scattering times in the range $6.5 \cdot 10^{-15}$-$2.6 \cdot 10^{-14}$ s for electron in Dirac cones and in the range $9.7 \cdot 10^{-15}$-$3.8 \cdot 10^{-14}$ s for holes in parabolic bands. Finally, Dirac cone Fermi velocities between 136 Km/s for LaFeAsO and 205 Km/s for SmFeAsO are calculated.

**Acknowledgements**

This work was carried out within the PRIN project No. 2008XWLWF9-005. FB acknowledges support from CASPUR under the Standard HPC Grant 2012. The authors acknowledge also the FP7 European project SUPER-IRON (grant agreement No. 283204).